\documentclass[%
 reprint,					
 amsmath,amssymb,
 superscriptaddress,
 aps,
 prd,
 longbibliography			
]{revtex4-1}

\usepackage{graphicx}
\usepackage{dcolumn}

\usepackage{hyperref}
\usepackage{xcolor}
\definecolor{darkblue}{rgb}{0.18,0.19,0.57}
\hypersetup{
    colorlinks,
    linkcolor=darkblue,
    citecolor=darkblue,
    urlcolor=darkblue
}

\usepackage{xspace}
\usepackage{braket}
\usepackage{color}
\usepackage[caption=false]{subfig}
\usepackage{siunitx}

\newcommand{\Exp}[1]{\ensuremath{\mathrm{e}^{#1}}\xspace}
\newcommand{\im}{\ensuremath{\mathrm{i}}\xspace}
\newcommand{\hc}{\ensuremath{\mathrm{H.c.}}\xspace}
\newcommand{\sign}[1]{\ensuremath{\mathrm{sgn}(#1)}\xspace}

\begin{document}

\title{Analog of cosmological particle creation in electromagnetic waveguides}

\author{Sascha Lang}
\email{s.lang@hzdr.de}
\affiliation{
 Helmholtz-Zentrum Dresden-Rossendorf, Bautzner Landstra{\ss}e 400, 01328 Dresden, Germany\\
}%
\affiliation{%
Fakult\"at f\"ur Physik, Universit\"at Duisburg-Essen, Lotharstra{\ss}e 1, 47057 Duisburg, Germany\\
}%
\author{Ralf Sch\"utzhold}
\affiliation{
 Helmholtz-Zentrum Dresden-Rossendorf, Bautzner Landstra{\ss}e 400, 01328 Dresden, Germany\\
}%
\affiliation{
 Institut f\"ur Theoretische Physik, Technische Universit\"at Dresden, 01602 Dresden, Germany
}%
\affiliation{%
Fakult\"at f\"ur Physik, Universit\"at Duisburg-Essen, Lotharstra{\ss}e 1, 47057 Duisburg, Germany\\
}%

\date{\today}

\begin{abstract}
We consider an electromagnetic waveguide with a time-dependent propagation speed $v(t)$
as an analog for cosmological particle creation.
In contrast to most previous studies which focus on the number of particles produced, 
we calculate the corresponding two-point correlation function. 
For a small step-like variation $\delta v(t)$, this correlator displays characteristic 
signatures of particle pair creation. 
As another potential advantage, this observable is of first order in the perturbation $\delta v(t)$,
whereas the particle number is second order in $\delta v(t)$ and thus 
stronger suppressed for small $\delta v(t)$. 
\end{abstract}

\pacs{Valid PACS appear here}

\maketitle

\section{\label{sec:introduction}Introduction}

Just a decade after Hubble's discovery of cosmic expansion \cite{Hubble_1929}, 
Schr\"odinger understood this mechanism to allow for 
particle creation out of the quantum vacuum \cite{schroedinger_1939}.
Being one of the most startling predictions of quantum field theory in curved 
space-times, cosmological particle creation was further studied by Parker \cite{Parker_1968} and others 
(see also \cite{BirrellDavies_1982}) in the late 1960's.
In the present universe, this effect is extremely tiny -- but, according to our 
standard model of cosmology, it played an important role for the creation of  
seeds of structure formation during cosmic inflation \cite{Mukhanov_1992}. 
Signatures of this process can still be observed today in the anisotropies of 
the cosmic micro-wave background radiation. 

As direct experimental tests of cosmological particle creation are probably 
out of reach, several laboratory analogs \cite{Unruh_1981, Visser_1998, Barcelo_2011} 
for quantum fields in expanding space-times have been proposed for various scenarios%
, including 
Bose-Einstein condensates \cite{Barcelo_2003, FedichevFischer_2003, Fischer_2004, 
FischerSchuetzhold_2004, Jain_2007, Prain_2010, Neuenhahn_2015, Eckel_2018}, 
ion traps  \cite{Alsing_2005, Schuetzhold_2007,Fey_2018,Wittemer_2019},  
and electromagnetic waveguides \cite{Laetheenmaeki_2013}. 
In the following, we shall consider the latter system (see also \cite{Tian_2017}),
which has already been used to observe an analog of the closely related dynamical 
Casimir effect 
\cite{Schuetzhold_1998, Johansson_2009, Johansson_2010, Wilson_2011, Laetheenmaeki_2013}.

Instead of the  often considered number of emerging particles, 
one can also study other observables, such as the two-point 
correlation function of the associated quantum field.
For condensed-matter analogs of black-holes 
(see, e.g., \cite{Unruh_1981, Visser_1998, SchuetzholdUnruh_2005, Barcelo_2011}), 
these correlations have already been studied in several works including 
\cite{Balbinot_2008, Carusotto_2008, MacherParentani_2009, SchuetzholdUnruh_2010}. 
In fact, the observation of analog Hawking radiation in Bose-Einstein 
condensates reported in \cite{Steinhauer_2016} was based on correlation measurements. 

Cosmological particle creation does also generate characteristic signatures
in the corresponding field correlations, see also \cite{Prain_2010}. 
Due to spatial homogeneity, particles are created in pairs with opposite momenta.
When both particles of a pair arrive at two suitable detectors at different space 
or space-time points $(t_1,x_1)$ and $(t_2, x_2)$, 
the associated signals are clearly correlated.

In the following, we study two-point correlations in an 
electromagnetic waveguide with a time-dependent effective speed of light $v(t)$.
Reducing this parameter $v(t)$ effectively increases all length scales in the 
set-up under consideration -- in analogy to cosmic expansion. 
Therefore, laboratory systems with a varying speed of light $v(t)$ ought to produce 
photon pairs in perfect analogy to the mechanism of cosmological particle creation.

\section{\label{sec:classicalModel}Classical waveguide model}

\begin{figure}
	\centering
	\includegraphics[width=0.95\linewidth]{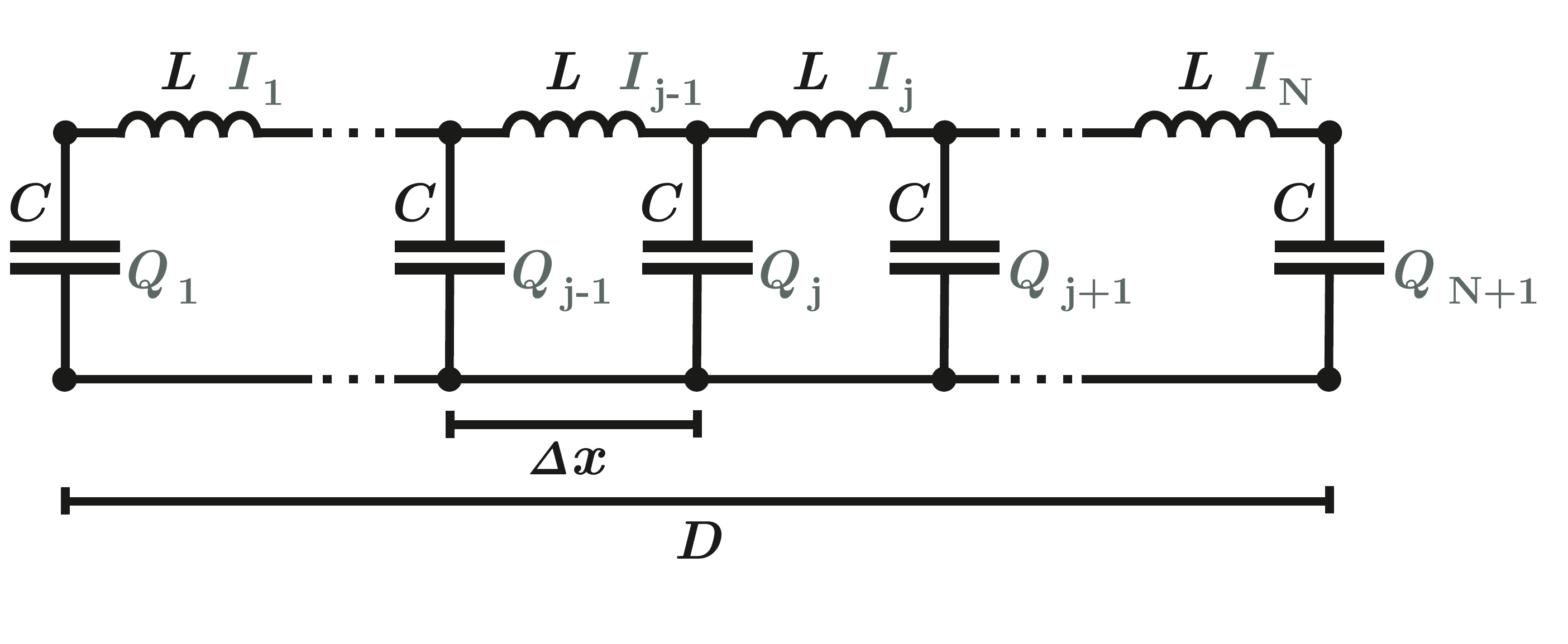}
	\caption{Illustration of an $LC$-circuit of length $D = N \, \Delta x$ that comprises $N$ discrete inductors  
					and $N + 1$ capacitors. The symbols $I_j$ and $Q_j$ denote the current in the $j$-th inductor and 
					the charge on the $j$-th capacitor respectively.}
	\label{fig:lcCircuit}
\end{figure}

Previous research on circuit quantum electrodynamics has brought up various possible 
implementations of waveguides with tunable parameters, see, e.g., 
\cite{Nation_2009, CastellanosBeltran_2008, Laetheenmaeki_2013, Tian_2017, Navez_2019}. 
Although realistic experiments are often based on superconducting quantum interference 
devices (SQUIDS), they typically still correspond to 
effective circuit diagrams. 
In this work, we will focus on waveguide structures that can be modeled with the 
effective set-up illustrated in Fig.~\ref{fig:lcCircuit}. 

Specifically, we consider an $LC$-circuit of total length $D$ which comprises $N + 1$ 
capacitors with equal capacities $C$ and $N$ inductors with equal but 
time-dependent inductances $L(t)$. 
Denoting the current in the $j$-th inductor with the symbol $I_j$ and the charge on the 
$j$-th capacitor with  $Q_j$, the Lagrangian of the set-up from Fig.~\ref{fig:lcCircuit} 
adopts the form
\begin{align}
\begin{aligned}
	\mathfrak{L}(t) = \sum_{j = 1}^{N+1}   \frac{1}{2 \;\! C} \; Q_j^2(t) - \sum_{j = 1}^{N} \frac{1}{2} \; L(t) \;\! I_j^2(t) \text{.}
	\label{eq:classicalDiscreteLagrangian}
\end{aligned}
\end{align}
Analogous to \footnote{`Supporting Information` to Ref. \cite{Laetheenmaeki_2013} 
\label{note:supportingInformation}}, we introduce a new generalized coordinate 
$\Phi_j(t)$ satisfying the relation \linebreak$\dot{\Phi}_j(t) = Q_j/\sqrt{C \; \Delta x}$. 
Apart from this, we eliminate all currents $I_j$ in the above Lagrangian $\mathfrak{L}(t)$ 
with the second line of the classical Kirchhoff's laws
\begin{align}
\begin{aligned}
	&\dot{Q}_{j} \!\!\! &=\text{\;}& I_{j} - I_{j-1}\,, \\
	& 0  \!\!\! &=\text{\;}& Q_{j}/C + \partial_t\left[ L(t) I_{j}\right] - Q_{j+1}/C\,\text{.}
	\label{eq:KirchhoffLaws}
\end{aligned}
\end{align}
Henceforth, we will further assume the length $\Delta x$ of each mesh in 
Fig.~\ref{fig:lcCircuit} to be significantly smaller than the characteristic 
wavelengths and the total length $D = N \Delta x$ of the waveguide. 
In the corresponding continuum limit of $\Delta x \rightarrow 0$ and $D = \text{const.}$, 
the Lagrangian $\mathfrak{L}(t)$ from equation \eqref{eq:classicalDiscreteLagrangian} 
turns into the expression 
\begin{align}
\begin{aligned}
	\mathfrak{L}(t) = \frac{1}{2} \int_{0}^{D} \!\!\! \mathrm{dx}   \left[ \left[ \dot{\Phi}(t,x)\right]^2 
							- v^2(t) \left[ \Phi^{\prime}(t,x)\right]^2  \right]\text{,}
	\label{eq:classicalLagrangian}
\end{aligned}
\end{align}
where the quantity $v(t) = \Delta x/\sqrt{L(t) \, C}$ accounts for the effective speed 
of light inside the circuit 
\footnote{
Note that later calculations for a sharp step-function $v(t)$ 
actually predict the creation of photons 
with arbitrary wavelengths $\lambda_n = 2 D/n, \text{ } n \in \mathbb{N}$. 
However, similar considerations for smoother profiles $v(t)$ suggest that particle 
production at large wave numbers $n$ is suppressed in real experiments 
\cite{BirrellDavies_1982}. 
Therefore, using the continuum limit is reasonably justified at least for the 
experimentally relevant modes satisfying $\lambda_n \gg \Delta x$. }.

Integrating the second line of equation \eqref{eq:KirchhoffLaws} with respect to $t$ 
further yields a relation $I(t,x) \propto \Phi^{\prime}(t,x)$ for the electric current 
in the continuum limit. 
As the waveguide depicted in Fig.~\ref{fig:lcCircuit} is isolated at both ends, 
the generalized flux $\Phi(t,x)$ inherits Neumann boundary conditions 
$\Phi^{\prime}(t,0) = \Phi^{\prime}(t,D) = 0$ $\forall \; t \in \mathbb{R}$ \cite{Note1}.

\section{\label{sec:canonicalQuantisation}Canonical quantization}

In order to quantize the classical model from above, we follow the path of canonical 
quantization and obtain the Hamiltonian 
\begin{align}
\begin{aligned}
	\hat{\mathfrak{H}}(t) = \frac{1}{2} \int_{0}^{D} \!\!\! \mathrm{dx} \left[ {\hat{\Pi}}^2(t,x)
										+  v^2(t) \left[ \hat{\Phi}^{\prime}(t,x)\right]^2  \right]
	\label{eq:quantumHamiltonian}
\end{aligned}
\end{align}
in which the operators $\hat{\Phi}(t,x)$ and $\hat{\Pi}(t,x)$ satisfy canonical 
commutation relations for a quantum field and its associated momentum.

The corresponding Heisenberg equations of motion can be combined to the wave equation 
\begin{align}
\begin{aligned}
	\ddot{\hat{\Phi}}(t,x) =\; v^2(t) \;\hat{\Phi}^{\prime \prime}(t,x)\text{.}
	\label{eq:operatorWaveEquation}
\end{aligned}
\end{align}
Bearing in mind that the field $\hat{\Phi}(t,x)$ has to satisfy Neumann boundary conditions, 
the mode functions
\begin{align}
\begin{aligned}
	&\Psi_{n=0}(x) \!\!\!& = & \,\, \sqrt{1/D} \\
 	&\Psi_{n>0}(x) \!\!\!& = & \,\, \sqrt{2/D} \cos{\left(\pi \, n \, x /D\right)}
 	\label{eq:completeSet}
\end{aligned}
\end{align}
allow for a decomposition 
\begin{align}
\begin{aligned}
 	\hat{\Phi}(t,x) = \sum_{n=0}^{\infty} \Psi_{n}(x) \; \hat{\varphi}_{n}(t)
 	\label{eq:phiOpAnsatz}
\end{aligned}
\end{align}
of the field operator $\hat{\Phi}(t,x)$, in which each term $\hat{\varphi}_{n}(t)$ 
constitutes a harmonic oscillator satisfying the differential equation
\begin{align}
\begin{aligned}
 	\ddot{\hat{\varphi}}_{n}(t) = - {\omega_{n}}^2(t) \; \hat{\varphi}_{n}(t)
			 \text{\quad with \quad }
	\omega_{n}(t) = \frac{\pi  n \, v(t)}{D}\,\text{.}
 	\label{eq:harmonicOscillator}
\end{aligned}
\end{align}

\section{\label{sec:stepLikeSpeedOfLight}Suddenly changing speed of light}

\subsection{\label{sec:stepLikeSpeedOfLight:operatorSolution}
Operator solution for a step-like profile $\boldsymbol{v(t)}$}

For a rapidly changing speed of light 
\begin{align}
\begin{aligned}
 	v(t) = \begin{cases}
 				v_0\text{,} & t < 0\\
 				v_1\text{,} & t > 0\text{,}
 			\end{cases}
 	\label{eq:stepwiseSpeedOfLight}
\end{aligned}
\end{align}
each operator $\hat{\varphi}_{n}(t)$ adopts a piecewise representation 
\begin{align}
\begin{aligned}
 	\hat{\varphi}_{n}(t<0) = \frac{1}{\sqrt{2 \,  \omega_{n}^{0}}} \left[\Exp{-\im \, \omega_{n}^{0}  t} \,\hat{a}_{n} 
																		+ \hc\right]\\
	\hat{\varphi}_{n}(t>0) = \frac{1}{\sqrt{2 \,  \omega_{n}^{1}}} \left[\Exp{-\im \, \omega_{n}^{1}  t} \,\hat{b}_{n} 
																		+ \hc\right]
 	\label{eq:harmonicOscillatorSolution}
\end{aligned}
\end{align}
with $\omega_{n}^{i} = \pi  n \, v_{i} /D$, where the expressions $\hat{a}_{n}$ 
and $\hat{b}_{n}$ satisfy canonical commutation relations for two separate sets of 
bosonic annihilators. 

The differential equation \eqref{eq:harmonicOscillator} requires each operator 
$\hat{\varphi}_{n}(t)$ and its temporal derivative $\partial_{t} \hat{\varphi}_{n}(t)$ 
to be continuous at $t = 0$, which implies the connection
\begin{align}
\begin{aligned}
 	\hat{b}_{n} = \frac{1}{2} \sqrt{\frac{v_{1}}{v_{0}}} \left[\left(1-\frac{v_{0}}{v_{1}}\right) 
								\hat{a}^{\dagger}_{n} + \left(1+\frac{v_{0}}{v_{1}}\right) \hat{a}_{n}\right]\text{.}
 	\label{eq:operatorsBeforeAndAfterConnection}
\end{aligned}
\end{align}

\subsection{\label{sec:stepLikeSpeedOfLight:particleCreation}Particle creation}

Based on the previous finding \eqref{eq:operatorsBeforeAndAfterConnection}, 
we can easily study how expectation values for the particle number operator 
\begin{align}
\begin{aligned}
 	\hat{N}_{n}(t) = \begin{cases} \hat{a}^{\dagger}_{n} \, \hat{a}_{n}\text{,}& t < 0\\[0.5em]
										\hat{b}^{\dagger}_{n} \, \hat{b}_{n}\text{,}& t > 0 
				  \end{cases}
 	\label{eq:particleNumberOp}
\end{aligned}
\end{align}
of the $n$-th mode evolve with time. 

In order to demonstrate the occurrence of particle production, we use the Heisenberg picture 
and study the expectation value $\bra{0} \hat{N}_{n}(t) \ket{0}$ for the initial vacuum 
state $\ket{0}$. 
Since this state satisfies the relation $\hat{a}_{n} \ket{0} = 0$ for all modes $n$, 
the number of photons inside the wave\-guide vanishes at all negative times. 
In the regime of $t > 0$, the particle number adopts the finite and constant value of 
$(v_1 - v_0)^2/(4 \, v_0 \, v_1)$ for all modes $n$ (see also \cite{Tian_2017}).

Consequently, particle creation for the sudden step $v(t)$ from 
equation \eqref{eq:stepwiseSpeedOfLight} occurs at the sharp instant of $t = 0$ 
and uniformly affects all modes 
\footnote{This finding implies the generation of photons even in modes with arbitrarily 
high $n$. 
However, as already pointed out in footnote \cite{Note2}, this finding just holds for 
sharp step-like profiles $v(t)$ and does not apply to more realistic set-ups with 
smoother parameter changes.}. 
However, the number of particles produced is of second order in the perturbation 
$\delta v = v_1 - v_0$ which might constitute a challenge for future experiments 
with small $\delta v$. 

\subsection{\label{sec:stepLikeSpeedOfLight:twoPointCorrelation}Two-point correlation for the operator $\boldsymbol{\hat{\Phi}(t,x)}$}

In case of a sharp step-function $v(t)$, the two-point correlation for the full field 
$\hat{\Phi}(t,x)$ can be evaluated analytically.  
In order to extract real results, we focus on the expression
\begin{align}
\begin{aligned}
 	\kappa(t_1,x_1,t_2,x_2)
						 =  \mathrm{Re}  \left[{\bra{0} \hat{\Phi}(t_1,x_1) \;\! \hat{\Phi}(t_2,x_2)  \ket{0}} \right] 
						 		\! - \chi_{\infty}
 	\label{eq:twoPointCorrelationDef}
\end{aligned}
\end{align}
which has been symmetrized with respect to an exchange of both space-time points $(t_1,x_1)$ 
and $(t_2,x_2)$. 
We further assume the term $\chi_{\infty}$ to compensate for the infinite but constant 
contribution of the infrared divergence associated with the lowest $(n=0)$-mode.

Exact analytic results for the correlation $\kappa(t_1,x_1,t_2,x_2)$ are calculated in 
Appendix \ref{appendix:correlationStepLike}. 
For all pairs of fixed times $t_1$ and $t_2$, the expression $\kappa(t_1,x_1,t_2,x_2)$ 
has logarithmic singularities along characteristic lines in the $(x_1,x_2)$-plane.

\subsubsection{\label{sec:stepLikeSpeedOfLight:twoPointCorrelation:negativeTimes}
Two-point correlation for negative $t_1$ and $t_2$}

If both times $t_1$ and $t_2$ are negative, the two-point correlation 
$\kappa(t_1\leq0,x_1,t_2\leq0,x_2)$ given in Appendix \ref{appendix:correlationStepLike} 
diverges to positive infinity under the condition 
\begin{align}
\begin{aligned}
 	x_1 + s_1 \, x_2 - s_2 \, v_0 \, (t_1 - t_2) = 2 \, D \,  m
 	\label{eq:singularitiesForBothTimesNegative}
\end{aligned}
\end{align}
with $s_1,s_2 \in \lbrace \pm 1\rbrace$ and $m \in \mathbb{Z}$.

\begin{figure}
	\centering
	\includegraphics[height=0.6\linewidth]{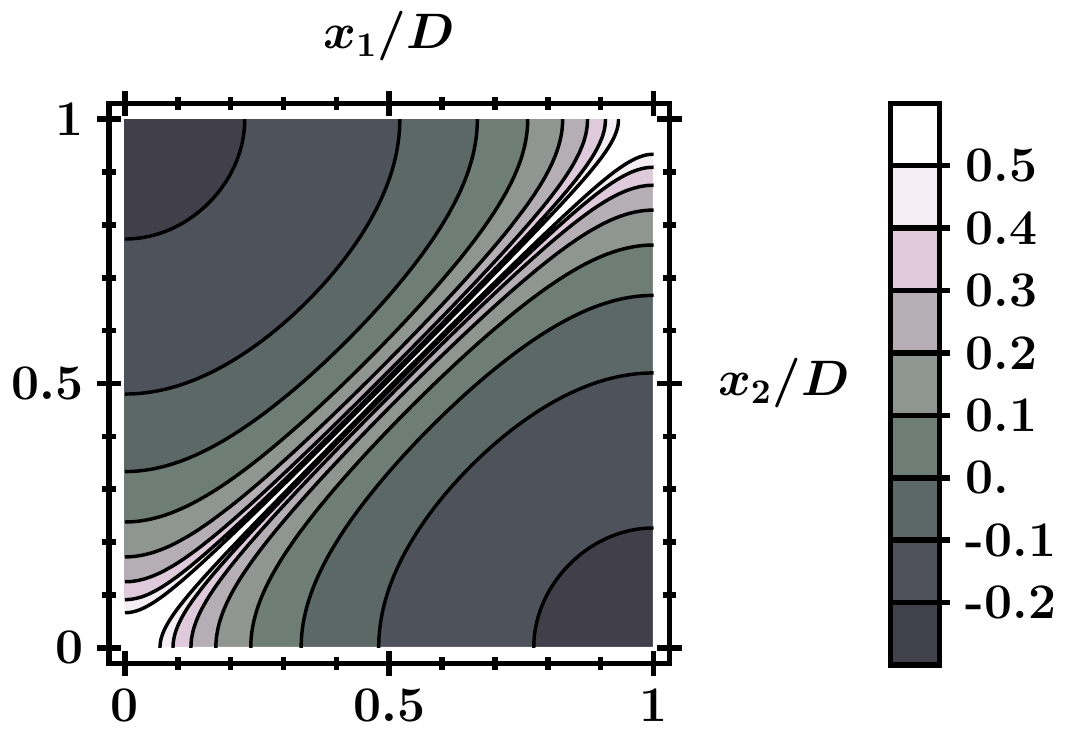}
	\caption{Rescaled equal-time correlation $v_0 \, \kappa(t_1,x_1,t_1,x_2)$ 
						plotted for an arbitrary argument $t_1 \leq 0$.
						Singularities just occur along the line with $x_1 = x_2$.}
	\label{fig:equalTimeCorrelationNegativeTimes}
\end{figure}

\begin{figure}
	\centering
	\includegraphics[width=0.5\linewidth]{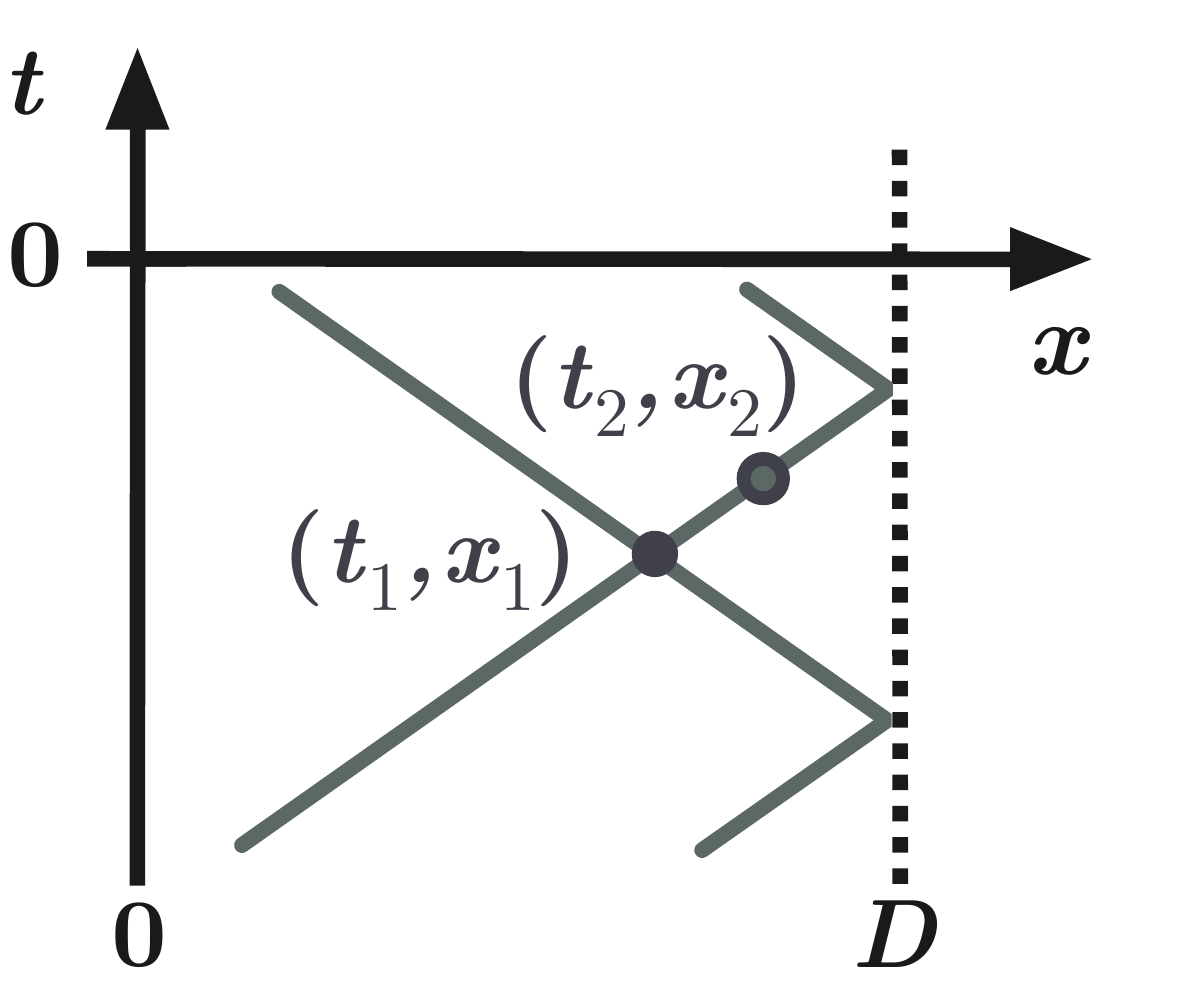}
	\caption{Worldlines for two different `signals` passing through a given space-time point $(t_1,x_1)$ at 
										velocities of absolute value $v_0$.  
										For all points $(t_2,x_2)$ on either worldline, the correlation function	
										 $\kappa(t_1\leq0,x_1,t_2\leq0,x_2)$ diverges.}
	\label{fig:illustrateLightConesForNegativeTimes}
\end{figure}

This identity just accounts for the standard light-cone singularities 
(possibly including reflections at the boundaries) in the initial vacuum state, 
see Figs.~\ref{fig:equalTimeCorrelationNegativeTimes}
and \ref{fig:illustrateLightConesForNegativeTimes}. 

\subsubsection{\label{sec:stepLikeSpeedOfLight:twoPointCorrelation:positiveTimes}
Two-point correlation for positive $t_1$ and $t_2$}

For positive times $t_1$ and $t_2$,  
the expression \linebreak $\kappa(t_1>0,x_1,t_2>0,x_2)$ 
calculated in Appendix \ref{appendix:correlationStepLike} adopts singularities under 
conditions of two different types 
\begin{align}
\begin{aligned}
 	&x_1 + s_1 \, x_2 - s_2 \, v_1 \, (t_1-t_2) = 2 \, D \,  m \\
 	&x_1 + s_1 \, x_2 - s_2 \, v_1 \, (t_1+t_2) = 2 \, D \,  m \text{.}
 	\label{eq:singularitiesForBothTimesPositive}
\end{aligned}
\end{align}
Except for a modified speed of light, the first identity from 
equation~\eqref{eq:singularitiesForBothTimesPositive} has the same form as the corresponding 
expression \eqref{eq:singularitiesForBothTimesNegative} for negative times $t_1$ and $t_2$. 
It thus also describes usual light-cone singularities.

\begin{figure}
	\centering
	\includegraphics[width=0.5\linewidth]{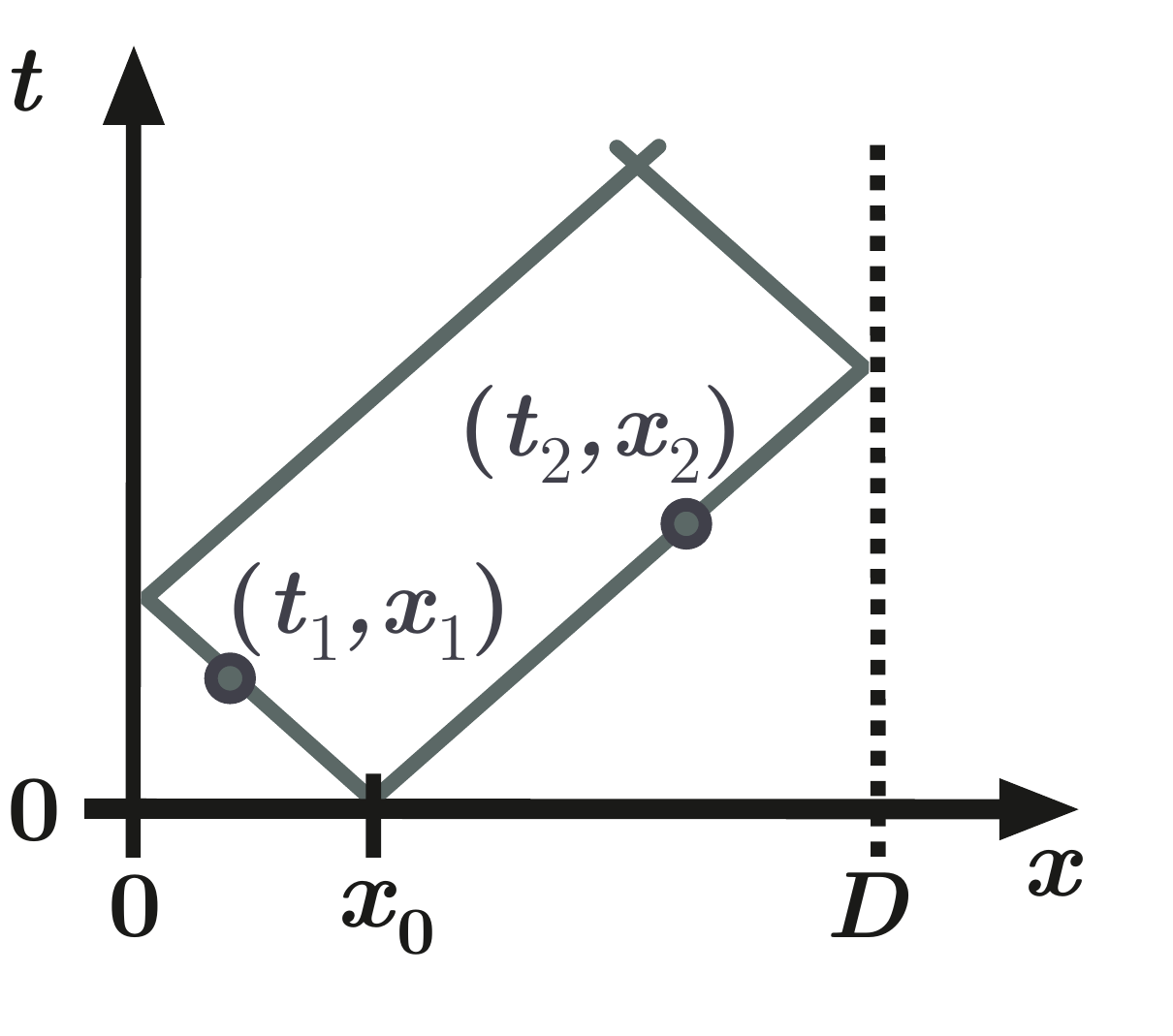}
	\caption{Worldlines for a photon pair which is produced at an arbitrary position $x_0$ and the sharp time $t = 0$. 
										For all points $(t_1,x_1)$ and  $(t_2,x_2)$ located on opposite worldlines, 
										the correlation function $\kappa(t_1>0,x_1,t_2>0,x_2)$ 
										has singularities characterized by the second line of equation 
										\eqref{eq:singularitiesForBothTimesPositive}.}
	\label{fig:illustrateLightConesForPositiveTimes}
\end{figure}

In contrast, the second type of singularities stems from the creation of particle pairs
at $t=0$. 
As one indication, the second line of equation \eqref{eq:singularitiesForBothTimesPositive}
is not invariant under time translation. 
As another indication, the corresponding pre-factors in the result 
$\kappa(t_1>0,x_1,t_2>0,x_2)$ from Appendix \ref{appendix:correlationStepLike} 
scale linearly with the perturbation $\delta v = v_1 - v_0$.
As an intuitive picture, one can imagine pair creation at the
sharp time $t=0$ and some random position $x_0$, where the produced particles propagate with 
opposite velocities $\pm v_1$. 
The condition $x_2 - x_1 = v_1 (t_1 + t_2)$ associated with this configuration is illustrated in Fig.~\ref{fig:illustrateLightConesForPositiveTimes}.

\begin{figure}
	\centering
	\subfloat{
		\includegraphics[height=0.6\linewidth]{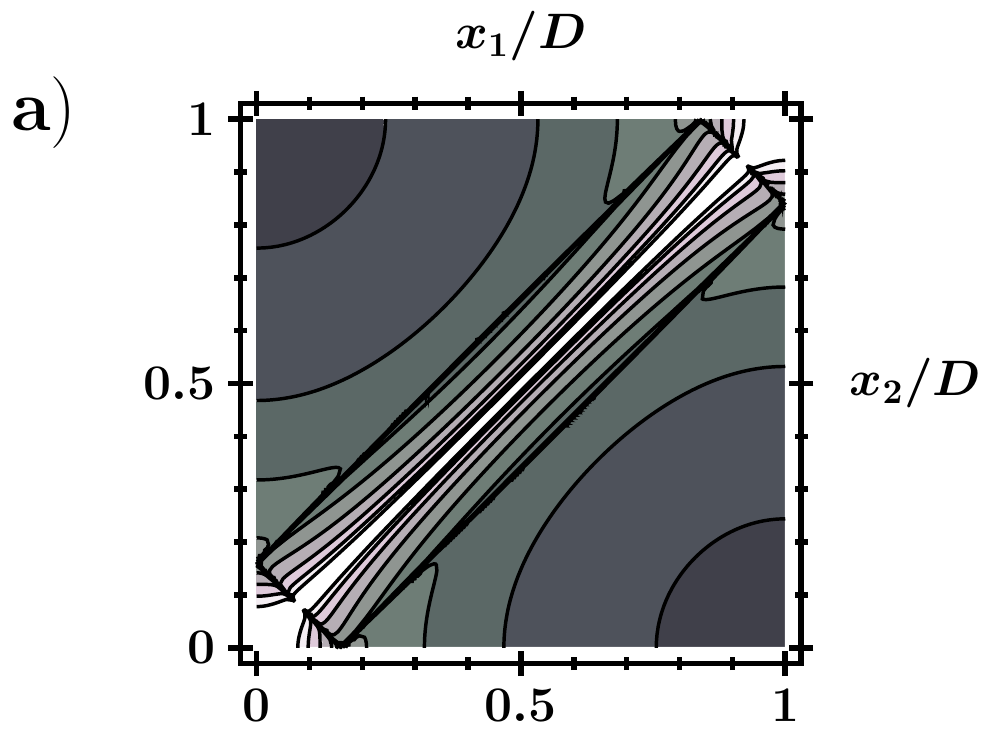}
		}\\
	\subfloat{
		\includegraphics[height=0.6\linewidth]{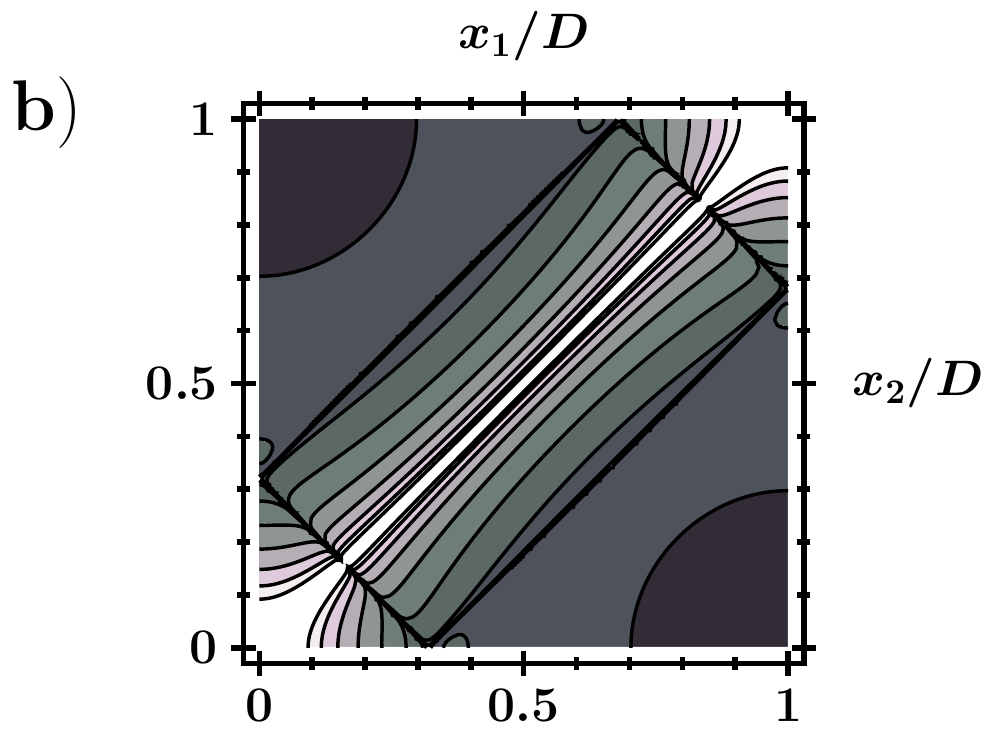}
	}
	\caption{Rescaled equal-time correlation $v_0 \, \kappa(t_1,x_1,t_1,x_2)$  
									plotted for two different arguments $t_1 > 0$ and velocities $v_0 = D$ as well as $v_1 = 0.8 \; D$. 
									The parameter $t_1$ adopts the value $0.1$ in plot a) and 
									$0.2$ in plot b). 
									\protect\newline
									Singularities described by the second line of equation \eqref{eq:singularitiesForBothTimesPositive} 
									occur along the black rectangular structure visible in both plots. 
									The colour scale is consistent with Fig.~\ref{fig:equalTimeCorrelationNegativeTimes}.}
	\label{fig:correlationResultBothTimesPositive}
\end{figure} 

Fig.~\ref{fig:correlationResultBothTimesPositive} provides plots of the correlation function 
$\kappa(t_1,x_1,t_2,x_2)$ for two fixed pairs of identical positive times $t_1 = t_2$. 
The singularities characterized by the second line of equation 
\eqref{eq:singularitiesForBothTimesPositive} occur along the black rectangle appearing 
in both of these plots. As a function of time, the corners of this structure continuously 
move along the boundaries of the domain $\left[0,D\right]^2$. 

Note that the sign of correlations along the rectangular pattern 
from Fig.~\ref{fig:correlationResultBothTimesPositive} depends on the ratio of both velocities 
$v_0$ and $v_1$. 
We obtain divergences to negative infinity if $v_0 > v_1$, and to positive infinity 
in case of $v_0 < v_1$. 
Thus, the former case $v_0 > v_1$ offers the advantage that singularities due to pair 
production are well-distinguishable from light-cone singularities. 

\subsubsection{\label{sec:stepLikeSpeedOfLight:twoPointCorrelation:oppositeTimes}
Two-point correlation for $t_1$ and $t_2$ having opposite signs}

If the times $t_1$ and $t_2$ differ in sign, we obtain singularities of the same types 
as in the previous paragraph. 
Given the exemplary case of $t_1 < 0$ and \linebreak $t_2 > 0$, 
divergences occur under the specific conditions 
\begin{align}
\begin{aligned}
 	x_1 + s_1 \, x_2 - s_2 \, (v_0 \, t_1 - &\, v_1 \, t_2) = 2 \, D \,  m \\
 	x_1 + s_1 \, x_2 - s_2 \, (v_0 \, t_1 + &\, v_1 \, t_2) = 2 \, D \,  m \text{.}
 	\label{eq:singularitiesForTimesWithOppositeSigns}
\end{aligned}
\end{align}
Taking the change of propagation velocity at $t = 0$ into account,
singularities specified by the first line of 
equation \eqref{eq:singularitiesForTimesWithOppositeSigns} can be associated with 
the light-cone once again.

\begin{figure}
	\centering
	\includegraphics[width=0.5\linewidth]{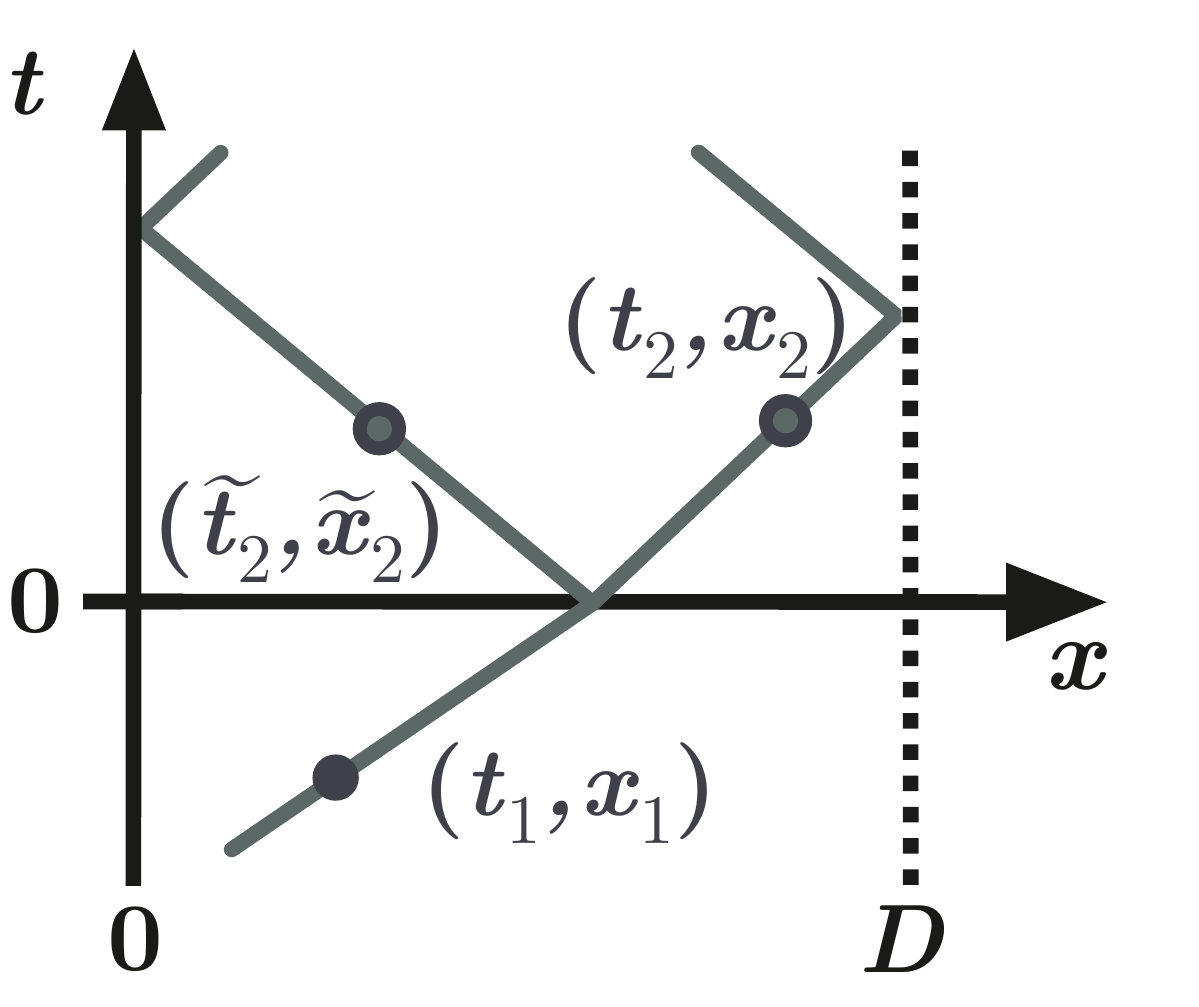}
	\caption{Two possible world-lines for an initial quantum fluctuation passing through a space-time point $(t_1,x_1)$ with $t_1 < 0$. 
						The condition in the first line of equation \eqref{eq:singularitiesForTimesWithOppositeSigns} 
						is satisfied for all points 
						$(t_2>0,x_2)$ that are located on the right branch of the depicted light-cone. 
						The second type of singularities emerges for those points 
						$(\tilde{t}_2>0,\tilde{x}_2)$ which belong to the other worldline illustrated above. 
						}
	\label{fig:worldlinesForTimesOfDifferentSigns}
\end{figure}

On the other hand, the second line of equation 
\eqref{eq:singularitiesForTimesWithOppositeSigns} indicates that quantum vacuum fluctuations 
propagating at an initial speed of either $+v_0$ or $-v_0$ are also partly reflected 
at the time $t=0$ and afterwards propagate with the new velocity $-v_1$ or $+v_1$ 
respectively. 
An illustration of both possible world-lines emerging from such a partial reflection 
is provided in {Fig.~\ref{fig:worldlinesForTimesOfDifferentSigns}}.
The splitting of initial fluctuations into superpositions of left- and right-moving 
components corresponds to the mixing of initial creation and annihilation operators 
$\hat{a}^{\dagger}_{n}$ and $\hat{a}_{n}$ in the new annihilators $\hat{b}_{n}$. 
Since this combination of terms $\hat{a}^{\dagger}_{n}$ and $\hat{a}_{n}$ 
is responsible for the occurrence of particle production, 
the partial reflections encoded in equation 
\eqref{eq:singularitiesForTimesWithOppositeSigns} illustrate 
that pair creation originates from fluctuations that have already been present 
at times $t < 0$.

\section{\label{sec:smoothStepLikeProfile}Continuously changing speed}

In order to assess whether the singularities obtained in Sect.~\ref{sec:stepLikeSpeedOfLight} 
also arise for smooth profiles $v(t)$, we repeat the previous calculations for a 
continuous function 
\begin{align}
\begin{aligned}
 	v^2(t) = \gamma_{-} \tanh{\!(t/\tau)} + \gamma_{+} \text{\quad with \quad \,} 
 	\gamma_\pm = \frac{v_1^2 \pm v_0^2}{2}\text{,}
 	\label{eq:tanhProfileVOfT}
\end{aligned}
\end{align}
where $\tau$ measures the finite time of change. 

Particle production in this modified set-up can be examined analogous to Sect. 3.4 of 
Ref.~\cite{BirrellDavies_1982} (see also \cite{Sauter_1932} for technical details). 
Again, we study the expression $\kappa(t_1,x_1,t_2,x_2)$ 
and analyze its behavior along the characteristic lines specified by 
equation \eqref{eq:singularitiesForBothTimesPositive}. 
This involves several approximations that are further discussed in  
Appendix \ref{appendix:smoothStepLikeProfile}. 

\subsection{\label{sec:smoothStepLikeProfile:operatorSolution}Operator solution for a smooth step $\boldsymbol{v(t)}$}

For the continuous profile $v^2(t)$ from equation \eqref{eq:tanhProfileVOfT}, 
solutions $\hat{\varphi}_n(t)$ of the differential equation \eqref{eq:harmonicOscillator} 
generally have non-trivial time dependencies. 
However, in the limiting cases of $t \rightarrow \pm \infty$, 
the operators $\hat{\varphi}_n(t)$ still adopt asymptotic representations 
equivalent to the result \eqref{eq:harmonicOscillatorSolution} for  a sharp step. 
Explicit calculations reveal the specific connection 
\begin{align}
\begin{aligned}
 	\hat{b}_n = \zeta_n^{(+)} \, \hat{a}_n  \,\, + \,\, \zeta_n^{(-)} \, \hat{a}^{\dagger}_n
 	\label{eq:aAndBOperatorConnectionTanhProfile}
\end{aligned}
\end{align}
with 
\begin{align}
\begin{aligned}
 	&\zeta_n^{(\pm)} \text{\!\!\!}&=\text{\;}& \sqrt{\frac{w^1_n}{w^0_n}}\,
 					 \frac{\Gamma\!\left[1\mp \im \,  w^0_n \tau \right] \, \Gamma\!\left[- \,\im \,  w^1_n \tau\right] }
 					 	  {\Gamma\!\left[1\mp\frac{\im }{2} \, (w^0_n \pm w^1_n) \, \tau\right] \, \Gamma\!\left[\mp\frac{\im}{2} \,
 					 	  (w^0_n \pm w^1_n) \, \tau\right]
 					 	  }
 	\label{eq:xsiExpressions}
\end{aligned}
\end{align}
between the asymptotic annihilator $\hat{b}_n$ after changing the speed of light 
and the corresponding initial ladder operators $\hat{a}_n$ and $\hat{a}_n^{\dagger}$, 
see also \cite{BirrellDavies_1982}.

\subsection{\label{sec:smoothStepLikeProfile:particleCreation}Particle creation for a smooth step $\boldsymbol{v(t)}$}

By evaluating expectation values $\bra{0} \hat{b}^{\dagger}_n \, \hat{b}_n \ket{0}$ 
for the initial vacuum  $\ket{0}$, we find the number of photons created in the $n$-th 
mode to adopt the well-known value \cite{Tian_2017, BirrellDavies_1982} 
\begin{align}
\begin{aligned}
 	\bra{0} \hat{b}^{\dagger}_n \, \hat{b}_n \ket{0} = \left\vert \zeta_n^{(-)} \right\vert^2 
 					= \frac{\sinh^2{\left[\frac{\pi}{2}\;\!(\omega_n^0-\omega_n^1)\;\!\tau\right]}}
 								{\sinh{\left[\pi\;\!\omega_n^0 \;\! \tau\right]}\sinh{\left[\pi \;\! \omega_n^1 \;\!\tau \right]}}
 	\label{eq:particleNumberSmoothStep}
\end{aligned}
\end{align}
for times $t \rightarrow \infty$ 
\footnote{The underlying calculations are based on the two identities 
${\left\vert \Gamma(\im y) \right\vert^2 = \pi / [y \, \sinh{(\pi y)}]}$ 
and ${\left\vert \Gamma(1+\im y) \right\vert^2 = \pi y /  \sinh{(\pi y)}}$ 
from  Eqs. (6.1.29) and (6.1.31) in Ref. \cite{AbramovitzStegun}.}.
For sharp step-functions with $\tau\to0$ or in case of $n = 0$, 
the above result reduces to the familiar expression $(v_0 - v_1)^2/(4 \, v_0 \, v_1)$. 
On the other hand, the  photon number $\bra{0} \hat{b}^{\dagger}_n \, \hat{b}_n \ket{0}$ 
undergoes exponential decay for $n \rightarrow \infty$, 
which results in a suppression of particle creation at short wavelengths 
$\lambda_n = 2 D / n$ 
\footnote{This finding resolves the issue raised in footnote \cite{Note2}.}.
Apart from this, particle production also vanishes if the continuous step $v(t)$ 
from equation \eqref{eq:tanhProfileVOfT} has a broad temporal width $\tau \rightarrow \infty$.

\subsection{\label{sec:smoothStepLikeProfile:correlationFunction}Two-point correlation after a smooth step $\boldsymbol{v(t)}$}

Unlike a sudden step-function $v(t)$, the continuous profile $v(t)$ from 
equation \eqref{eq:tanhProfileVOfT} does not yield a compact expression 
for the two-point correlation $\kappa(t_1,x_1,t_2,x_2)$. 
However, for sufficiently large times $t_1$ and $t_2 \gg \tau$, 
further discussions in Appendix \ref{appendix:smoothStepLikeProfile} 
provide an approximate result $\kappa(t_1,x_1,t_2,x_2)$ 
that correctly includes the contributions of modes with large $n$.
Since all singularities of the quantity $\kappa(t_1,x_1,t_2,x_2)$ arise from 
high modes approaching $n \rightarrow \infty$,  the large-$n$ approximation 
in Appendix \ref{appendix:smoothStepLikeProfile} clearly reveals whether 
a smooth step $v(t)$ yields the same divergent contributions as its 
discontinuous counterpart from equation \eqref{eq:stepwiseSpeedOfLight}.

As expected, we find the two-point correlation $\kappa(t_1,x_1,t_2,x_2)$ to 
still diverge under the light-cone conditions 
\begin{align}
\begin{aligned}
 	x_1 + s_1 \, x_2 - s_2 \, v_1 \, (t_1-t_2) = 2 \, D \,  m \text{,}
 	\label{eq:singularitiesForBothTimesPositiveSmoothA}
\end{aligned}
\end{align}
while the additional singularities due to pair creation 
\begin{align}
\begin{aligned}
 	x_1 + s_1 \, x_2 - s_2 \, v_1 \, (t_1+t_2) = 2 \, D \,  m 
 	\label{eq:singularitiesForBothTimesPositiveSmoothB}
\end{aligned}
\end{align}
from Sect. \ref{sec:stepLikeSpeedOfLight} are smoothened out for continuous profiles $v(t)$.
This can be explained by the fact that particle creation can no longer be 
associated with a sharp point of time.

\section{\label{sec:conclusion}Conclusion}

As a laboratory analog for cosmological particle creation, we have considered a 
waveguide with a time-dependent speed of light $v(t)$ and calculated the 
two-point correlation $\kappa(t_1,x_1,t_2,x_2)$ for the generalized flux variable 
$\hat{\Phi}(t,x)$.  
First, we studied a sudden step function $v(t)$.
In addition to the usual light-cone singularities (possibly including 
reflections at the boundaries), we found a distinctive pattern of logarithmic 
singularities in $\kappa(t_1,x_1,t_2,x_2)$ which clearly reflects the 
dynamics of pair creation occurring at a sharp instant of time. 
If we replace the sudden step in $v(t)$ by a smooth profile,
those additional singularities are smoothened out.  
Nevertheless, the correlation $\kappa(t_1,x_1,t_2,x_2)$ displays 
distinctive signatures of pair creation, which could be observed 
experimentally. 

In contrast to the number of particles produced (see, e.g.,~\cite{Tian_2017}), 
the imprint of pair creation onto the correlation function 
$\kappa(t_1,x_1,t_2,x_2)$ is of first order in the perturbation 
$\delta v = v_1 - v_0$. 
Therefore, we propose that measuring two-point correlations instead of 
particle numbers may enhance the chances for observing analog 
cosmological particle creation in future experiments with tunable waveguides.

\acknowledgements 

R.S.~acknowledges support by DFG (German Research Foundation), grant 278162697 (SFB 1242).

\appendix
\section{\label{appendix:correlationStepLike}Calculation of the symmetrized two-point correlation for a rapid step $\boldsymbol{v(t)}$}
In order to work out the symmetrized two-point correlation $\kappa(t_1,x_1,t_2,x_2)$ for a step-like profile $v(t)$, we insert the findings \eqref{eq:phiOpAnsatz} and \eqref{eq:harmonicOscillatorSolution} into equation \eqref{eq:twoPointCorrelationDef}. 
After evaluating all quantum mechanical expectation values, elementary trigonometric identities can be used to rearrange the function $\kappa(t_1,x_1,t_2,x_2)$ into a sum containing multiple expressions of the characteristic shape
\begin{align}
\begin{aligned}
 	\sum_{n=1}^{\infty} \frac{p^n  \cos{\left(n\, \xi \right)}}{n} = -\frac{1}{2}\ln{\left[1-2 \, p\cos{\left(\xi\right)} + p^2\right]} 
 	 \text{,}&&	p^2\leq1\text{,}			
 	\label{eq:usefulSum}
\end{aligned}
\end{align}
with here $p = 1$, where the result on the right-hand side has been taken from Eq. (1.448) in Ref. \cite{GradshetynRyzhik_1965}. 

Typical arguments $\xi$ occurring in terms of the specific shape \eqref{eq:usefulSum} can be abbreviated with a symbol
\begin{align}
\begin{aligned}
 	\xi^{(v_i,v_j \vert \pm)}_{s_1,s_2}&(t_1,x_1,t_2,x_2) \\
								&= \frac{\pi}{D}\left[x_1 + s_1 \, x_2 - s_2 \,  (v_i \, t_1 \pm v_j \, t_2)\right]
 	\label{eq:usefulXiTerm}
\end{aligned}
\end{align}
in which the indices $s_1$ and $s_2 \in \lbrace \pm1 \rbrace$ constitute placeholders for two variable signs.

By applying the previous considerations to expressions  $\kappa(t_1,x_1,t_2,x_2)$ with different combinations of signs $\sign{t_1}$ and $\sign{t_2}$, we obtain the specific results
\begin{align}
\begin{aligned}
 	&\kappa(t_1\leq0,x_1,t_2\leq0,x_2)  = \! - \frac{1}{8\pi \, v_0} \,    \\
						&\text{\qquad\qquad} \times  \sum_{s_i=\pm 1}  \ln{\!\left[ 2 - 2 \cos{\!\left[\xi^{(v_0, v_0\vert-)}_{s_1,s_2}
															(t_1,x_1,t_2,x_2)\right]}\right]}\text{,}
 	\label{eq:correlationResultBothTimesNegative}
\end{aligned}
\end{align}
\begin{align}
\begin{aligned}
 	&\kappa(t_1>0,x_1,t_2>0,x_2) = \! -\frac{1}{16\pi \, v_0} \sum_{\gamma = \pm 1} 
										\left[1- \gamma \, \frac{{v_0}^2}{{v_1}^2}\right] \\
							&\text{\qquad\qquad \;\;} \times \sum_{s_i = \pm 1} \ln{\!\left[ 2 - 2 \cos{\!\left[
												\xi^{(v_1, v_1\vert\gamma)}_{s_1,s_2}(t_1,x_1,t_2,x_2)\right]}\right]}
 	\label{eq:correlationResultBothTimesPositive}
\end{aligned}
\end{align}
and
\begin{align}
\begin{aligned}
 	&\kappa(t_1\leq0,x_1,t_2>0,x_2) = \! -\frac{1}{16\pi \, v_0} \sum_{\gamma = \pm 1} 
										\left[1- \gamma \, \frac{{v_0}}{{v_1}}\right] \\
							&\text{\qquad\qquad \;\;} \times \sum_{s_i = \pm 1} \ln{\!\left[ 2 - 2 \cos{\!\left[
												\xi^{(v_0, v_1\vert\gamma)}_{s_1,s_2}(t_1,x_1,t_2,x_2)\right]}\right]}\text{.}
 	\label{eq:correlationResultForTimesOfDifferentSigns}
\end{aligned}
\end{align}

For arbitrary fixed arguments $t_1$ and $t_2$, the above findings \eqref{eq:correlationResultBothTimesNegative} to \eqref{eq:correlationResultForTimesOfDifferentSigns} have logarithmic singularities along characteristic lines in the $(x_1,x_2)$-plane. More specifically, such singularities arise if the respective term $\xi^{(v_0, v_0\vert-)}_{s_1,s_2}(t_1,x_1,t_2,x_2)$, $\xi^{(v_1, v_1\vert\gamma)}_{s_1,s_2}(t_1,x_1,t_2,x_2)$ or $\xi^{(v_0, v_1\vert\gamma)}_{s_1,s_2}(t_1,x_1,t_2,x_2)$ corresponds to an integer multiple of $2\pi$.

\section{\label{appendix:smoothStepLikeProfile}Approximate result for the symmetrized two-point correlation after a smooth step $\boldsymbol{v(t)}$}
\subsubsection{\label{appendix:smoothStepLikeProfile:correlationFunction:genStructure}General structure of the two-point correlation for large times $t_1$, $t_2 \gg \tau$}
For times $t \gg \tau$ located after the smooth step $v(t)$ from Sect. \ref{sec:smoothStepLikeProfile}, the full quantum field $\hat{\Phi}(t,x)$ adopts the asymptotic representation
\begin{align}
\begin{aligned}
 	&\hat{\Phi}(t \gg \tau, x) 
 	\text{\!\!\!}&=\;& \sum_{n=0}^{\infty} \frac{ \Psi_n(x)}{\sqrt{2 \,  \omega_{n}^{1}}} 
 												\left[\Exp{-\im \, \omega_{n}^{1}  t} \,\hat{b}_{n} + \hc\right]
 	\label{eq:fullFieldOperatorSmoothStep}
\end{aligned}
\end{align}
in which the terms $\hat{b}_n$ are given by equation \eqref{eq:aAndBOperatorConnectionTanhProfile}. 

After inserting the above finding into the symmetrized correlation function \eqref{eq:twoPointCorrelationDef}, the identities already given in footnote \cite{Note4} can be used to extract the result 
\begin{align}
\begin{aligned}
 	\kappa(t_1,x_1,t_2,x_2) = \kappa_\mathrm{A}(t_1,x_1,t_2,x_2) + \kappa_\mathrm{B}(t_1,x_1,t_2,x_2)
 	\label{eq:correlationSmoothStepA}
\end{aligned}
\end{align}
with 
\begin{align}
\begin{aligned}
 	&\kappa_\mathrm{A}
 					&\text{\!\!\!}= \; &
 							\sum_{n=1}^{\infty} \frac{1}{2 \, \omega_n^1} \, \Psi_n(x_1) \, \Psi_n(x_2)  \,
 									\cos{\left[ \omega_n^1 (t_1 - t_2) \right]} \\
 					&&\times\text{\,}&	\frac{	 \sinh^2{\!\left[\frac{\pi}{2} (\omega_n^0 + \omega_n^1) \, \tau\right]}
 										 	 +   \sinh^2{\!\left[\frac{\pi}{2} (\omega_n^0 - \omega_n^1) \, \tau\right]}
 											 }
 											 {   \sinh{\!\left[\pi \, \omega_n^0 \,\tau\right]}
 											 	 \sinh{\!\left[\pi \, \omega_n^1 \,\tau\right]}
 											 } 									
 	\label{eq:correlationSmoothStepKappaA}
\end{aligned}
\end{align}
and 
\begin{align}
\begin{aligned}
 	&\kappa_\mathrm{B}	
 					=  \sum_{n=1}^{\infty} \frac{4 \pi}{\left[(\omega_n^0)^2-(\omega_n^1)^2\right] \tau} \, \Psi_n(x_1) \, \Psi_n(x_2) \\
 	&\text{\quad}\times\, \mathrm{Re}\Bigg[
 								\frac{\Exp{-\im \, \omega_n^1 \, (t_1 + t_2)}}{\sinh{\!\left(\pi \;\!\omega_n^0 \;\!\tau\right)}} 
 								\left( \frac{	\Gamma{\!\left[- \im \, \omega_n^1 \, \tau \right]}		}
 										 		{   \Gamma{\!\left[-\frac{\im \left(\omega_n^0 + \omega_n^1\right) \tau }{2} \right]}
 										 	 		\Gamma{\!\left[ \frac{\im \left(\omega_n^0 - \omega_n^1\right) \tau }{2} \right]}
 								} \right)^2	
 						  \Bigg]\text{.}					
 	\label{eq:correlationSmoothStepKappaB}
\end{aligned}
\end{align}

\subsubsection{\label{appendix:smoothStepLikeProfile:correlationFunction:approixmations}Expansions allowing for an explicit evaluation}
As both expressions $\kappa_\mathrm{A}$ and $\kappa_\mathrm{B}$ involve sums that lack straightforward analytic solutions, the following studies rely on approximations of the respective summands. In order to assess whether the two-point correlation $\kappa(t_1,x_1,t_2,x_2) $ after a smooth step $v(t)$ acquires the same singularities as the corresponding expression \eqref{eq:correlationResultBothTimesPositive} for a rapidly changing speed of light, it is sufficient to examine the contributions of modes with large indices $n \rightarrow \infty$.

Based on the asymptotic expansion 
\begin{align}
\begin{aligned}
 	\sinh{\pi x} \sim \Exp{\pi x}/2 \text{\qquad for } x \rightarrow \infty	
 	\label{eq:sinhAsymptotics}
\end{aligned}
\end{align}
and the Stirling formula 
\begin{align}
\begin{aligned}
 	\Gamma(z) \sim \sqrt{2\pi} \, \Exp{-z} \, z^{z-1/2} \text{\qquad for } \vert z \vert \rightarrow \infty	\text{,}
 	\label{eq:stirlingApprox}
\end{aligned}
\end{align}
we can significantly simplify both expressions \eqref{eq:correlationSmoothStepKappaA} and \eqref{eq:correlationSmoothStepKappaB}. 

Numerical studies reveal that the relative error associated with the approximation \eqref{eq:sinhAsymptotics} is negligibly small for all arguments $x \geq 1$. Apart from this, the Stirling formula \eqref{eq:stirlingApprox} also yields at least qualitatively reliable results for all  purely imaginary arguments $z = \im  x$ with $\vert x \vert > 1$. 

Bearing in mind the relation $\omega_{n}^{i} = \pi  n \, v_{i} /D$, the expansions \eqref{eq:sinhAsymptotics} and \eqref{eq:stirlingApprox}  are clearly applicable to all summands in the expressions $\kappa_\mathrm{A}$ and $\kappa_\mathrm{B}$  that have sufficiently large indices $n$. Moreover, they even hold for smaller integers $n \gtrsim 1$ if the step-width $\tau$ of the continuous profile $v(t)$ exceeds the characteristic times $D/(\pi v_0)$, $D/(\pi v_1)$ and $2 D/\vert \pi (v_0 - v_1)\vert$. 

For the exemplary set-up studied in Ref. \cite{Laetheenmaeki_2013}, the values $D = \SI{4}{mm}$ and $v_0 = 0.5 \, c_0$ constitute realistic experimental parameters with $c_0$ denoting the vacuum speed of light. If we further assume $v_1 = 0.45 \, c_0$, the asymptotic expansions \eqref{eq:sinhAsymptotics}  and \eqref{eq:stirlingApprox} hold for all indices $n \in \mathbb{N}$ as long as $\tau \geq \SI{1.7e-10}{\second}$.

In the opposite case of $\tau$ violating the above requirements, the subsequent results contain incorrect contributions for modes with small indices $n$. Nevertheless, any conclusions concerning the appearance of singularities remain valid even if the underlying approximations are unreliable for low integers $n$.

\subsubsection{\label{sec:smoothStepLikeProfile:correlationFunction:termKappaA}Approximate result for the term $\kappa_\mathrm{A}(t_1,x_1,t_2,x_2) $}
By applying the asymptotic expansion \eqref{eq:sinhAsymptotics} to each summand of equation \eqref{eq:correlationSmoothStepKappaA}, we obtain the approximate result 
\begin{align}
\begin{aligned}
 	&\kappa_\mathrm{A} &\text{\!\!\!}\approx \; & 
 							\sum_{n=1}^{\infty} \frac{\Psi_n(x_1) \, \Psi_n(x_2)}{2 \;\! \omega_n^1} \,
 							 \cos{\left[ \omega_n^1 \, (t_1 - t_2) \right]} \\
 	&&&						 	\text{\qquad\qquad\qquad\;\;} \times \left[1 + 
 										\Exp{- 2 \pi \min{\left\lbrace\omega_n^0,\omega_n^1\right\rbrace} \, \tau}
 										\right]							
 	\label{eq:kappaAApproxStepOne}
\end{aligned}
\end{align}
that can be further simplified analogous to Appendix \ref{appendix:correlationStepLike}. 

More specifically, we expand the square brackets in the last term of equation \eqref{eq:kappaAApproxStepOne}, use the identity \eqref{eq:usefulSum} with $p_1 = 1$ or $p_2 = \exp{\left[-2 \, \pi^2 \, \min{\left\lbrace v_0,v_1 \right\rbrace} \, \tau /D)\right]}$ respectively and finally retain the expression 
\begin{align}
\begin{aligned}
 	&\kappa_\mathrm{A}   
 						&\text{\!\!\!}\approx \; & - \frac{1}{8\pi \, v_1} \; \sum_{k=1,2} \; \sum_{s_i=\pm 1} \\
 						&&& 	\times  \ln{\!\left[ 1 - 2 \, p_k \cos{\!\left[\xi^{(v_1, v_1\vert-)}_{s_1,s_2}
															(t_1,x_1,t_2,x_2)\right]} + p_k^2 \right]}\text{.}
 	\label{eq:correlationResultBothTimesPositiveKappaA}
\end{aligned}
\end{align}
The $(k \! = \! 1)$-contribution to the latter result obviously resembles the $(\gamma \! = \! -1)$-terms in the corresponding function $\kappa(t_1>0,x_1,t_2>0,x_2)$ from Appendix \ref{appendix:correlationStepLike}. Therefore, the expression $\kappa_\mathrm{A}(t_1,x_1,t_2,x_2)$ similarly adopts singularities under the condition
\begin{align}
\begin{aligned}
 	x_1 + s_1 \, x_2 - s_2 \, v_1 \, (t_1-t_2) = 2 \, D \,  m\text{.}
 	\label{eq:singularitiesForBothTimesPositiveSmoothAAppendix}
\end{aligned}
\end{align} 

On the other hand, all terms satisfying $k \! = \! 2$ remain finite for arbitrary combinations of space-time points $(t_1, x_1)$ and $(t_2,x_2)$.

\subsubsection{\label{sec:smoothStepLikeProfile:correlationFunction:termKappaB}Approximate result for the term $\kappa_\mathrm{B}(t_1,x_1,t_2,x_2) $}
Based on the Stirling formula \eqref{eq:stirlingApprox}, the squared parentheses in equation \eqref{eq:correlationSmoothStepKappaB} can be approximated according to
\begin{align}
\begin{aligned}
 	&\left( \frac{	\Gamma{\!\left[- \im \, \omega_n^1 \, \tau \right]}		}
 				{   \Gamma{\!\left[-\frac{\im \left(\omega_n^0 + \omega_n^1\right) \tau }{2} \right]}
 					\Gamma{\!\left[ \frac{\im \left(\omega_n^0 - \omega_n^1\right) \tau }{2} \right]}
 					} \right)^2\\	
 	&\text{\qquad\quad}\approx \text{\,}
 	\frac{\im \, n \, \tau}{8 \, D} \frac{\left( v_0^2-v_1^2 \right)}{v_1} 
 			\Exp{\frac{\pi^2 \, \tau}{D} \, n \, (v_0 - v_1) \,  \Theta(v_0 - v_1)}\\
 	&\text{\qquad\qquad\quad}\times	
 				\Exp{-\im \frac{\pi \, n \, \tau}{D} \left[ v_0  \ln{\left(\frac{\left\vert v_0 - v_1 \right\vert}{v_0 + v_1} \right)}
 						 +   v_1  \ln{\left(\frac{4 v_1^2}{\left\vert v_0^2 - v_1^2 \right\vert}\right)}\right]}\text{,}
 	\label{eq:gammaTermApproximation}
\end{aligned}
\end{align}
where the symbol $\Theta$ denotes the Heaviside-function.

If we likewise replace the factor $\sinh{\!\left(\pi \, \omega_n^0 \tau\right)}$ by means of equation \eqref{eq:sinhAsymptotics}, the resulting expression $\kappa_\mathrm{B}$ can be further reduced to the form
\begin{align}
\begin{aligned}
 	\kappa_\mathrm{B} \,	
 					&\approx  
 					\sum_{s_i \in \left\lbrace \pm 1\right\rbrace} \sum_{n=1}^{\infty} -\frac{s_2}{2 \,  \pi \, v_1 }  \\
 	&\times\, 
 					\frac{\Exp{-\frac{\pi^2 n \, \tau}{D} \min{\left\lbrace v_0, v_1\right\rbrace}}}{n} 
 					\sin{\left[ n \, \tilde{\xi}_{s_1,s_2}(t_1,x_1,t_2,x_2) \right]}							
 	\label{eq:correlationSmoothStepKappaBPartTwo}
\end{aligned}
\end{align}
with 
\begin{align}
\begin{aligned}
 	&\tilde{\xi}_{s_1,s_2}(t_1,x_1,t_2,x_2) 
								= \frac{\pi}{D}
								\bigg[
									x_1 + s_1 \, x_2 - s_2 \, v_1 \, (t_1 + t_2)\\
	&\text{\qquad \qquad }		- s_2 \, v_0 \, \tau \, \ln{\left(\frac{\left\vert v_0 - v_1 \right\vert}{v_0 + v_1} \right)}
 						 			- s_2 \, v_1 \, \tau \, \ln{\left(\frac{4 v_1^2}{\left\vert v_0^2 - v_1^2 \right\vert}\right)}
								\bigg]\text{.}					
 	\label{eq:xiTildeDefinition}
\end{aligned}
\end{align}

By afterwards using the identity 
\begin{align}
\begin{aligned}
 	\sum_{n=1}^{\infty} \frac{p^n \, \sin{(n x)}}{n} = \arctan{\left(\frac{p \sin{x}}{1- p \cos{x}}\right)} 
 	\text{,} &&	p^2 \leq 1	\text{,}	
 	\label{eq:sumOverSinTerms}
\end{aligned}
\end{align}
taken from Eq. (1.448) in Ref. \cite{GradshetynRyzhik_1965}, we finally obtain the approximate result
\begin{align}
\begin{aligned}
 	&\kappa_\mathrm{B}	  
 					\approx  
 					\sum_{s_i \in \left\lbrace \pm 1\right\rbrace}  -\frac{s_2}{2 \pi \, v_1 }  \\
 	&\text{\quad}\times\, 
 					\arctan{\left(	\frac{	\sin{[\tilde{\xi}_{s_1,s_2}(t_1,x_1,t_2,x_2)]} 	}
 										{	\Exp{\frac{\pi^2 \,\tau}{D} \min{\left\lbrace v_0, v_1\right\rbrace}}
 												- 	\cos{[\tilde{\xi}_{s_1,s_2}(t_1,x_1,t_2,x_2)]}
 										}
 							\right)}\text{.}							
 	\label{eq:correlationSmoothStepKappaBPartThree}
\end{aligned}
\end{align}

Since the arctangent adopts finite values for all real arguments, the expression $\kappa_\mathrm{B}$ never diverges. This finding requires all singularities obeying the second line of equation \eqref{eq:singularitiesForBothTimesPositive} to be smoothened out for continuous profiles $v(t)$.

In the limiting case of a broad step $v(t)$ meeting the requirements $\tau \geq D/(\pi v_0)$ and $\tau \geq D/(\pi v_1)$, the term $\exp{\left[\frac{\pi^2 \, \tau}{D} \min{\left\lbrace v_0, v_1\right\rbrace}\right]}$ adopts constant values significantly greater than $1$. The space-time dependency of each summand in equation \eqref{eq:correlationSmoothStepKappaBPartThree} is thus mainly determined by the sine-like numerator and undergoes a change of sign along approximately those lines satisfying the condition $\tilde{\xi}_{s_1,s_2}(t_1,x_1,t_2,x_2) = m \, \pi$ with $m \in \mathbb{Z}$.

For weak perturbations $\delta v = v_1 - v_0$, the last two terms of equation \eqref{eq:xiTildeDefinition} reduce to a small offset and the function $\tilde{\xi}_{s_1,s_2}(t_1,x_1,t_2,x_2)$ hence approaches the corresponding expression $\xi^{(v_1, v_1\vert +)}_{s_1,s_2}(t_1,x_1,t_2,x_2)$ from Appendix \ref{appendix:correlationStepLike}. As a result, the separate summands in equation \eqref{eq:correlationSmoothStepKappaBPartThree} change their signs under approximately the condition
\begin{align}
\begin{aligned}
 	x_1 + s_1 \, x_2 - s_2 \, v_1 \, (t_1+t_2) =  D \,  m
 	\label{eq:changeOfSignForBothTimesPositiveSmoothBAppendix}
\end{aligned}
\end{align} 
with $m \in \mathbb{Z}$. 

For even values of $m$, the latter identity reproduces the characteristic lines that are associated with pair production emerging from the corresponding rapid step $v(t)$. Even if the associated singularities do not persist for smoother profiles $v(t)$, the expression $\kappa_\mathrm{B}$ still has a distinctive pattern in a similar parameter regime. 

After combining all separate summands to the full quantity $\kappa_\mathrm{B}$, the resulting expression \eqref{eq:correlationSmoothStepKappaBPartThree} acquires broad steps under the specific conditions provided in the second line of equation \eqref{eq:singularitiesForBothTimesPositive}.


\bibliographystyle{apsrev4-1} 

\end{document}